\documentclass[]{spie}
\usepackage[]{graphicx}
\usepackage{subcaption}
\usepackage{wrapfig}
\usepackage[numbers]{natbib}
\usepackage{float}
\usepackage{multicol}
\usepackage{comment}

\usepackage[pdftex,
    pdfauthor={Daniel Vagg, Derek O'Callaghan, Fionn O hOgain, Sheila McBreen, Lorraine Hanlon, David Lynn, and William O'Mullane},
    pdftitle={GAVIP: A Platform for Gaia Data Analysis},
    pdfsubject={GAVIP, PaaS, Gaia, collaborative, data processing, platform},
    pdfcreator={},
    pdfproducer={}
]{hyperref}

\title{GAVIP: A Platform for Gaia Data Analysis}

\author{Daniel Vagg\supit{a}, Derek O'Callaghan\supit{a}, Fionn {\'O} h{\'O}g{\'a}in\supit{a}, Sheila McBreen\supit{a,b}, Lorraine
Hanlon\supit{a,b}, David Lynn\supit{b}, and William O'Mullane\supit{c}
\skiplinehalf
\supit{a}Parameter Space, University College Dublin, Belfield, Dublin, Ireland; \\
\supit{b}School of Physics, Science Centre North, University College Dublin, Belfield, Dublin, Ireland; \\
\supit{c}European Space Astronomy Centre, Madrid, Spain
}

\authorinfo{Further author information: (Send correspondence to Parameter Space Ltd.)\\Parameter Space Ltd: E-mail: info@parameterspace.ie}
 
\begin{document} 
\maketitle 

\begin{abstract}
Gaia is a major European Space Agency (ESA) astrophysics mission designed to map and analyse 10$^9$ stars, ultimately generating more than 1~PetaByte of data-products. 
As Gaia data becomes publicly available and reaches a wider audience, there is an increasing need to facilitate the further use of Gaia products without needing to download large datasets. 
The Gaia Added Value Interface Platform (GAVIP) is designed to address this challenge by providing an innovative platform within which
scientists can submit and deploy code, packaged as ``Added Value Interfaces'' (AVIs), which will be executed close to the data.
Deployed AVIs and associated outputs may also be made available to other GAVIP platform users, thus providing a mechanism for scientific
experiment reproducibility.
This paper describes the capabilities and features of GAVIP.
\end{abstract}


\section{INTRODUCTION} 
\label{sec:introduction}
    There is currently more than 1 petabyte (PB\footnote{1 PB = 1,000 TB}) of astronomy data that is accessible  electronically, with archives  growing at a rate of 0.5 PB per year. 
Data rates and sensor sizes are increasing as more new facilities come online, leading to predictions that by 2020 more than 60 PB of archived data will be accessible. 
The existence of these massive, distributed, and heterogeneous data-sets poses challenges, both to the underlying data curation and
infrastructure/archive management and to traditional astronomical research methods.
A major challenge for future astronomy missions is that the on-board instruments are so sensitive and have such large fields of view that the volume of data is transcending current working capabilities. 
Traditionally, astronomers would obtain observations of a source, these data would be downloaded and analysed, and although the analysis can be very complex and may need to be processed on dedicated servers, the data were not prohibitively large as to prevent users downloading products and working locally.
This is changing with large scale surveys such as Gaia \citep{2001A&A...369..339P},  Intermediate Palomar Transient Factory (iPTF)
\citep{2009PASP..121.1395L}, Pan-STARRS \citep{2010SPIE.7733E..0EK},  and  the Large Synoptic Survey Telescope (LSST)
\citep{ivezic2008lsst}.
These ground and space-based instruments with wide fields of view and/or scanning capabilities now observe the sky continuously, build up all-sky maps regularly and have opened a huge discovery space.
Within this context, if the right tools are provided, more and diverse users will be motivated to take advantage of the science within these growing archives.

Gaia is an extremely ambitious and complex ESA mission designed to carry out an all-sky astrometric, photometric and spectroscopic survey of objects brighter than 20 magnitude for astrometric and photometric observations, and 16 magnitude for spectroscopic observations \citep{2001A&A...369..339P}. 
It was launched in 2013 and is currently orbiting the second Lagrange point at a distance of 1.5 million kilometres from the Earth in the anti-Sun direction.
Gaia will measure of order one 10$^9$ objects, including about 1\% of the stars in the Milky Way, 10$^{6}$ to 10$^{7}$ galaxies, 500,000 quasars \citep{Gaia}, and of order 6000 supernova \citep{2003MNRAS.341..569B,LuriOverviewCatalogueGOG2014,RobinGUMS2012}. 
The Gaia scanning law gives the number of times a region will be re-observed over its 5-year mission, and comes from this spinning motion of
the satellite and its orbit around the Sun. The average number of observations per object is 70, although it can be as low as a few tens or as high as 200~\cite{2012Ap&SS.341..163A}.
During the mission, Gaia is expected to transmit some 150 terabytes (TB) of raw data to Earth, leading to production of a catalogue of
10$^{9}$ individual objects.
After on-ground processing, the full database is expected to be in the range of one to two PB of data~\citep{LuriOverviewCatalogueGOG2014}.

\begin{figure}[!t]
    \centering
    \includegraphics[width=0.8\textwidth]{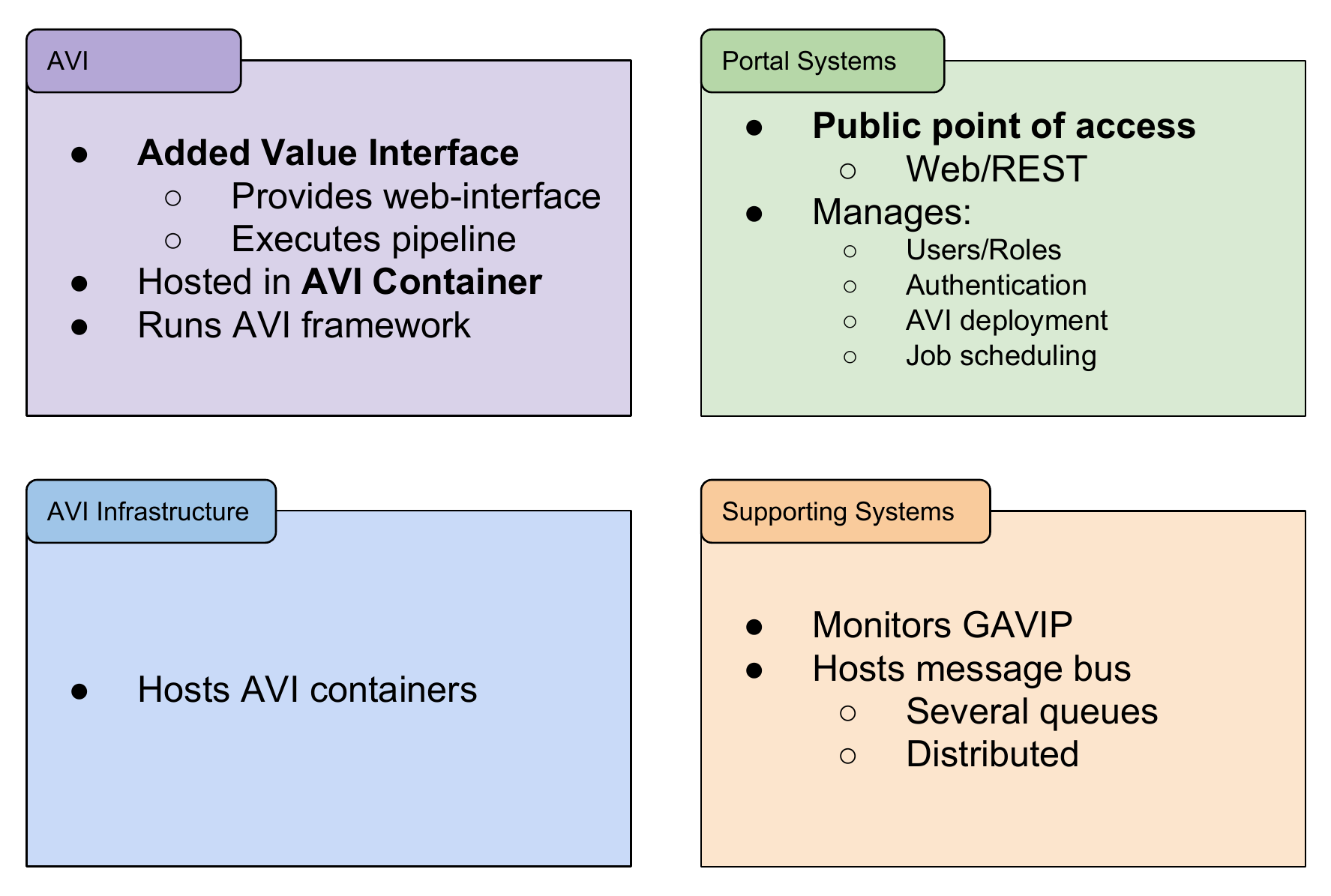}
    \caption{High-level elements of the GAVIP platform.}
    \label{fig:highlevel_components}
\end{figure}

\begin{figure}[!t]
    \centering
    \includegraphics[width=0.8\textwidth]{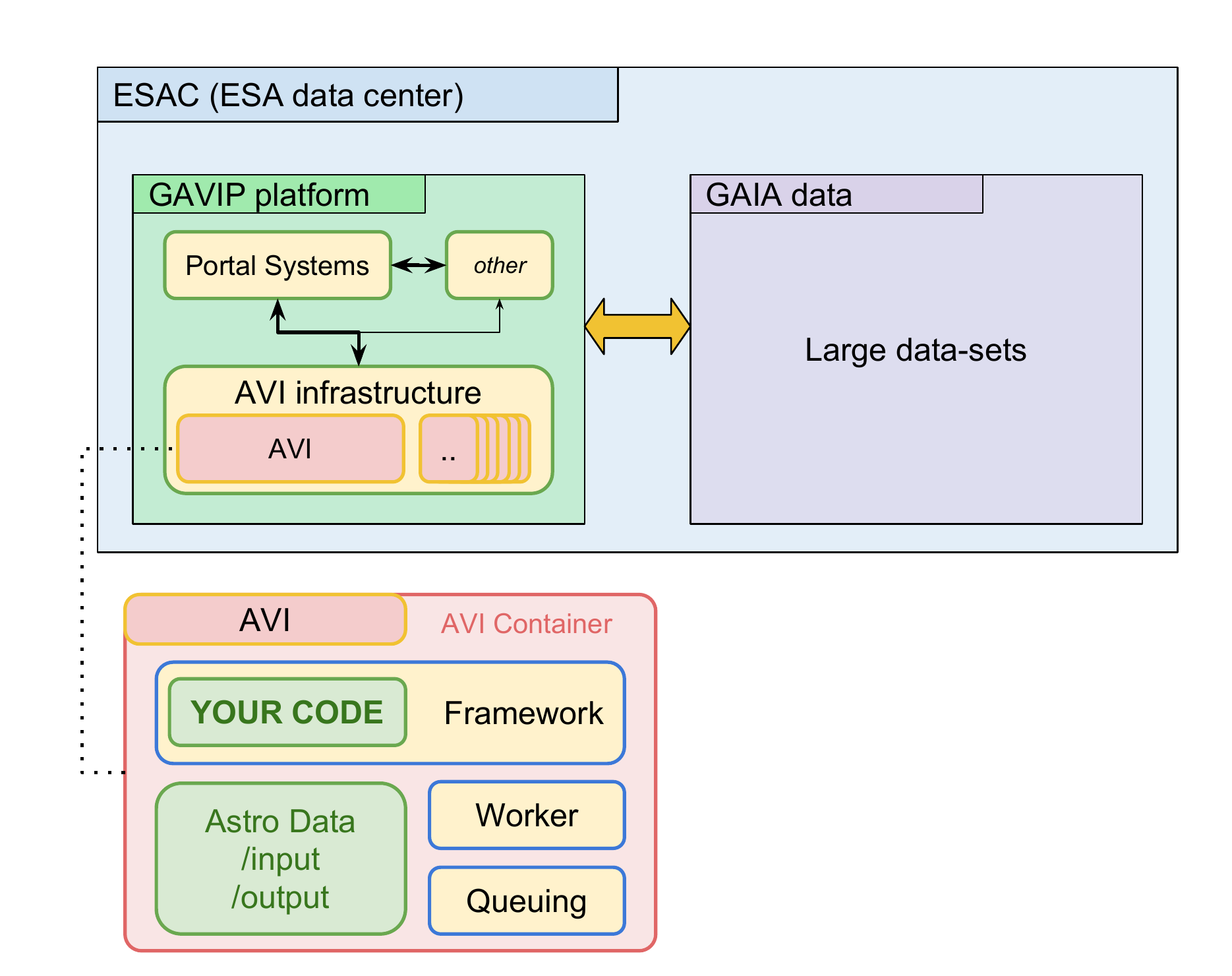}
    \caption{GAVIP platform overview illustrating interrelationship between the high-level elements of Figure~\ref{fig:highlevel_components}.}
    \label{fig:platform_overview}
\end{figure}

Consequently, the large volume of data that will ultimately be generated by Gaia means that alternatives to the traditional approach of
local data analysis should be investigated. Other more general issues to consider include those related to the reproducibility of scientific experiments, amid a growing concern about irreproducible
results\footnote{\url{http://www.nature.com/news/reproducibility-1.17552}}.
The Gaia Added Value Interface Platform (GAVIP) is a Python-based platform designed
to address these challenges by providing a mechanism for reproducible analysis of the Gaia archive data.
There are two primary objectives of GAVIP. 
The first objective is to allow user-contributed algorithms, packaged as Added Value Interfaces (AVIs), to be executed near the
Gaia archive so that data can be quickly acquired over internal network infrastructure.
For example, it should be possible to package asteroid shape modelling code written in Fortran as simply as a stellar lightcurve classifier written in Java.
The second objective is to allow users to easily use these AVIs, and share their results.
The platform conceptually consists of four high-level elements: multiple AVIs; portal systems which
provide the interface to the platform; AVI infrastructure to host AVIs; supporting systems.
These elements are illustrated in Figure~\ref{fig:highlevel_components}.
The interrelationship of these elements, and a simplified view of the internal structure of an AVI can be found in
Figure~\ref{fig:platform_overview}.

There are a wide range of requirements and diverse challenges involved in delivering such a platform, including interface, flexibility, performance and scalability requirements.
Some of the most important requirements have been refined and provided in Table~\ref{tab:gavip_requirements}.
A more detailed discussion of challenges in building the GAVIP platform is provided in this paper, setting the context for a detailed technical view of GAVIP and how it is designed to address these challenges. 
Related work is discussed in Section~\ref{sec:related_work}, providing an overview of related platforms and introducing one of the first
design choices of GAVIP (virtualisation using Docker containers).
The GAVIP concept is then presented in Section~\ref{sec:gavip_concept}, including an overview of GAVIP operations and the key
features of the platform.
Next, Section~\ref{sec:architecture} describes the architecture and technical implementation of the GAVIP platform.
Finally, Section~\ref{sec:conclusions} contains the overall conclusions and offers suggestions for future work.

\begin{table}[H]
    \centering
    \begin{tabular}{|p{4cm}|p{12cm}|}
        \hline
        \textbf{Requirement} & \textbf{Description} \\ \hline
        \hline
        Automation & The GAVIP platform must be highly automated, requiring little to no operator intervention. 
        \\ \hline
        Web-based & GAVIP users shall be able to completely interact with the platform and its AVIs using a web-browser. 
        \\ \hline
        AVI flexibility & Each AVI must be highly configurable by the user, including the arbitrary selection of input parameters. 
        \\ \hline
        AVI isolation & Following the GAVIP automation requirement, AVI submission must be automated, leading to the removal of manual code
        inspection. Individual AVI design may result in undesired behaviour when handling concurrent requests from multiple users. Therefore, AVIs must
        be isolated as far as possible. 
        \\ \hline 
        Data sharing & GAVIP shall facilitate across-AVI sharing of both inputs and outputs. 
        \\ \hline 
        Simultaneous usage & GAVIP must support the simultaneous execution of multiple AVIs (including multiple instances of the same AVI). 
        \\ \hline 
        Scalability & As the GAVIP community increases, and AVI pipelines become more complex, the platform must be able to scale in order
        to support growing demand. 
        \\ \hline 
        Load control & GAVIP must provide functionality to prevent the capacity of the system being exceeded. 
        \\ \hline
    \end{tabular}
    \caption{GAVIP Requirements.}
    \label{tab:gavip_requirements}
\end{table}


\section{RELATED WORK} 
\label{sec:related_work}
    In this section, related work to the GAVIP platform is discussed, focusing on similar platforms and services.
Prior to this discussion, a brief introduction to the technical requirements of virtualisation and large scale data analysis
is provided for context, as virtualisation is one of the primary techniques used to isolate software.

\subsection{Virtualisation Methods} 
\label{sub:virtualisation_methods}
\begin{figure}[H]
    \begin{subfigure}{.5\textwidth}
        \centering
        \includegraphics[width=5cm]{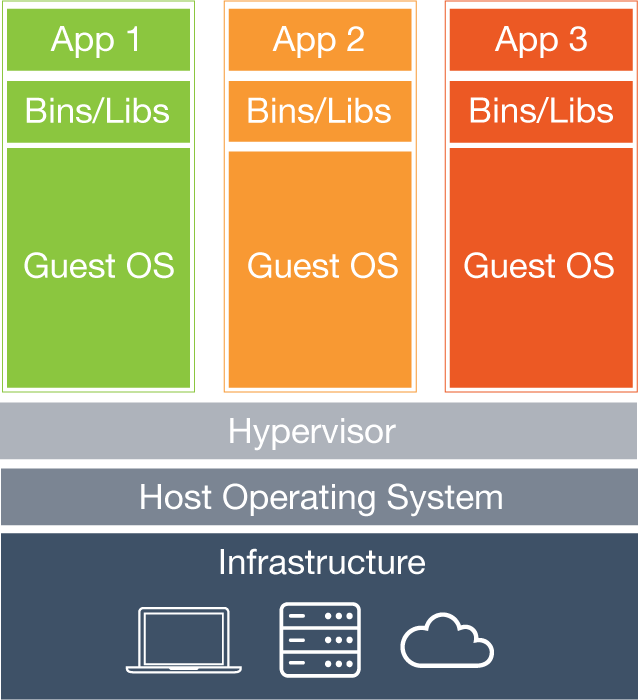}
        \caption{Virtual Machines}
    \end{subfigure}%
    \begin{subfigure}{.5\textwidth}
        \centering
        \includegraphics[width=5cm]{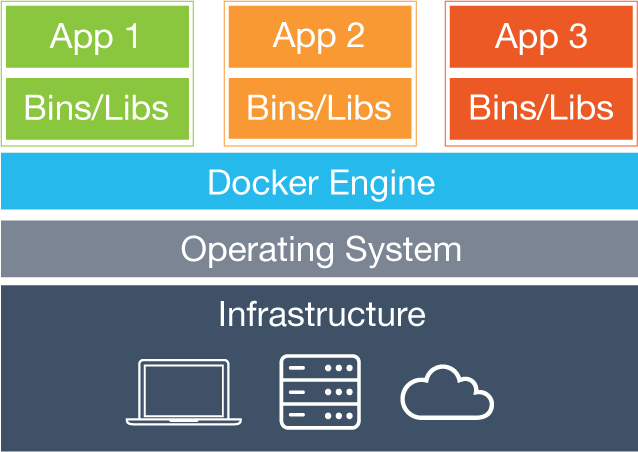}
        \caption{Containers}
    \end{subfigure}
    \caption{Comparative illustration of virtual machine and container virtualisation showing the use of a hypervisor and multiple guest
    Operating Systems when creating virtual machines. \textbf{Source:} Docker documentation \cite{DockerImages}}
    \label{fig:vms_vs_containers}
    \end{figure}

    One primary GAVIP requirement is AVI isolation (see Table~\ref{tab:gavip_requirements}). Two options were considered to achieve this,
    namely (1) Virtual Machines, and (2) operating system (OS) level virtualisation (containers). The latter was
    selected, implemented using Docker~\citep{Merkel:2014:DLL:2600239.2600241}, principally due to the differences in resource usage between
    the two techniques. 
    In the case of virtual machines, a hypervisor is used to emulate computer hardware which an operating system, libraries, and application must be installed on. 
    However, with containers, a virtual environment is provided that shares the majority of the underlying
    operating system; this is illustrated in Figure~\ref{fig:vms_vs_containers}.
    In addition to the reduced RAM footprint with container virtualisation, there is also an improvement in hardware utilization as
    the hardware (for example, the CPU or RAM) is directly accessed rather than being used through an emulator (the hypervisor).

    Beyond the benefits in resource usage and utilisation, containers can be created and started much more quickly than virtual machines
    (often in under a second), as they share the majority of the host operating system.
    This gives GAVIP the ability to create and start AVIs on demand, for individual users. 
    This ability to run containers on demand is used to support running AVIs in different modes of operation, which is one of the key techniques used to manage platform resources (see section \ref{ssub:resource_management}).
    Containers also offer a unique level of application portability. 
    This is especially useful when sharing applications within a community, as a container image can be copied and used to provide an
    exact copy of a software environment.
    In the case of GAVIP, it is crucial that this is the case, as astronomy software can rely on a wide array of packages, for example,
    NumPy~\citep{numpy2011} and SciPy~\citep{scipy2001} that require compilation for the environment in which they run.

\subsection{Existing data analysis tools} 
\label{sub:existing_tools}
    A number of platforms have recently emerged that address some of the contemporary data analysis requirements. For example, the Google
    Cloud Datalab~\citep{GoogleDatalab} provides a facility for hosting and running analysis tools on large data-sets.
    The challenge of general reproducibility of analysis and scientific experiments is a motivation for Binder~\citep{MyBinder}.
    Both of these platforms use a combination of OS virtualisation (for example, Docker) and Jupyter Notebooks\citep{Jupyter} as part of their
    underlying technology. 
    Within the astronomy community, there are a number of a platforms which have been built to facilitate exploitation of astronomy data;
    for example, the Canadian Astronomy Data Centre (CADC - ~\citealp{CADC}), and SciServer~\cite{SciServer}.
    Both of these platforms offer users the ability to interact with large data sets, by running analysis within the platform's infrastructure.
    The CADC platform provides a portal for several data-products, aswell as services to interact with these data-products. While most of the services are online tools for data inspection, a ``Virtual Machine on Demand'' service is provided which allows users to create batch processing jobs against data-archives. 
    SciServer is a modular platform comprising of multiple components which allow users to perform complex queries on large databases, inspect the Sloan Digital Sky Survey, store data products, cross match astronomical datasets, and access SciServer services using Python, R, and Matlab scripts~\cite{SciServerTools}.
    Both of these platforms have similarities with the GAVIP platform; they both allow users to perform remote analysis and store their data-products. 
    However, their approaches to post processing of data-products are different.
    The SciServer platform provides an extensible range of tools, but they are developed within the platform, and can't be provided by external users. 
    CADC allows users to provide their own analysis tool (through ``Virtual Machine on Demand''), but this is purely for executing a custom processing task.
    GAVIP provides a combination of these features, allowing users to provide their own tools (comprised of both processing jobs and interfaces), and easily share them within the platform for other users.

    GAVIP uses Docker as its underlying container engine, and involves Jupyter as part of its recommended AVI development practices, similar to both Binder and Google Datalab. 
    However, GAVIP also encourages the evolution of algorithms, often captured in notebooks, to form interactive web-tools packaged as AVIs. 
    This design aids exposure of the data and its analysis as AVIs to the wider scientific community, without the need for programming expertise. 
    The GAVIP concept is discussed further in Section~\ref{sec:gavip_concept}, with the underlying architecture presented in
    Section~\ref{sec:architecture}.


\section{GAVIP CONCEPT} 
\label{sec:gavip_concept}
    As illustrated in Figure~\ref{fig:highlevel_components}, GAVIP consists of a portal, AVIs, AVI infrastructure, and supporting systems used by the platform.
The GAVIP portal provides the web interface to the platform, and is the central point of access for all platform users. 
The portal allows users to submit, browse, and execute AVIs, as well as inspect resulting data products.
AVIs consist of a Docker container (referred to as the AVI container) running user-contributed code within the AVI framework. 
The user-contributed code, referred to as the AVI code, generally consists of a web interface and one or more pipelines. 
A pipeline is a set of tasks within an AVI that perform some level of analysis or further processing of the Gaia data archive, ultimately
providing added value.
Pipelines and interfaces of AVIs are decoupled within GAVIP to more cleanly separate the resource demands within the platform, as a web
interface typically requires fewer resources in comparison to pipeline algorithms.
Further detail on the platform architecture is provided in Section~\ref{sec:architecture}.

\subsection{GAVIP User Roles} 
\label{sub:usage_scenarios}
    GAVIP defines four types of roles for registered users. 
    Users may have more than one defined role.
    An additional role is used to define behaviour for anonymous users.
    Each of these roles are designed to satisfy different platform usage; their primary purposes are briefly described in
    Table~\ref{tab:gavip_user_roles}.
    
    \begin{table}[H]
        \centering
        \begin{tabular}{|p{4cm}|p{12cm}|}
            \hline
            \textbf{GAVIP Role} & \textbf{Description} \\ \hline
            \hline
            Operator & Maintains and configures the system, including viewing and managing all container deployments as required.  
            \\ \hline
            Developer & Creates and submits AVIs to the GAVIP platform. Discussion of AVI development scenarios are provided later in section
            \ref{ssub:impl_avi_development}. 
            \\ \hline 
            Scientist & Browses the AVI catalogue within GAVIP, starts and uses AVIs. 
            \\ \hline
            Outreach user & Can browse and interact with AVIs similar to Scientists, but is offered alternative interfaces for simpler
            interaction with the archive, for the sake of outreach. 
            \\ \hline 
            Anonymous user & Can browse and interact with AVIs similar to Scientists, but is restricted by the amount of resources that
            may be consumed in doing so. 
            \\ \hline
        \end{tabular}
        \caption{GAVIP User Roles.}
        \label{tab:gavip_user_roles}
    \end{table}

\subsection{GAVIP Features} 
\label{sub:gavip_features}
    An overview of the high-level features in GAVIP can be found in Table~\ref{tab:gavip_features}. These features have been derived from
    the requirements in Table~\ref{tab:gavip_requirements}.
    They are further described in more technical detail in Section~\ref{sec:architecture}.

    \begin{table}[H]
        \centering
        \begin{tabular}{|p{4cm}|p{12cm}|}
            \hline
            \textbf{Feature} & \textbf{Overview} \\ \hline
            \hline
            AVI framework & The AVI framework provides the foundation for AVIs and is used convert an algorithm into a reusable and
            shareable tool. 
            \\ \hline 
            User-space & The user-space is a persistent storage volume that is accessible by all AVIs. The platform includes a browser for
            the user-space. 
            \\ \hline 
            Data-sharing & To minimize unnecessary processing of data, users can share data products. Other users can reuse these
            data products in their AVIs. 
            \\ \hline 
            Dashboard & A dashboard is provided in GAVIP to enable simultaneous interaction with multiple AVI interfaces. 
            \\ \hline 
            Command Line Interface (CLI) & A CLI to the platform is provided by the GAVIP client. 
            \\ \hline 
            AVI Development & AVI development is the process of creating an AVI outside of the platform in ``standalone'' mode, and 
            submitting it to the platform when ready. This process has been designed to be relatively simple, non-restrictive, while also ensuring compataibility with the GAVIP platform. See section \ref{ssub:avi_development} for more detail.  \\ \hline
        \end{tabular}
        \caption{GAVIP features}
        \label{tab:gavip_features}
    \end{table}

    GAVIP is designed for use on a data archive which is itself scheduled for full release years after the platform is completed. 
    One of the driving challenges of GAVIP is providing a platform that will not become obsolete before the data archive is fully available. 
    AVIs are ultimately what will attract users to the GAVIP platform; the value of the platform from the scientific community's perspective is provided only by AVIs.
    Three exemplar AVIs are currently under development, and will be deployed in GAVIP prior to public release of the full archive. 
    However, the GAVIP ecosystem must expand for the platform to reach its full potential.
    This introduces the challenges associated with the AVI framework (\ref{ssub:the_avi_framework}). 
    Beyond the design of the framework, there are also challenges in how AVIs are developed (\ref{ssub:avi_development}). 
    Once AVIs are functional and simple to develop, management of AVI resources leads to the majority of the challenges in GAVIP (\ref{ssub:resource_management}).

    \subsubsection{The AVI framework} 
    \label{ssub:the_avi_framework}
        The AVI framework is embedded within AVI containers, and provides the infrastructure and much of the boilerplate functionality required for an AVI.
        This framework must be:
        \begin{enumerate}
            \item Able to support the queueing and execution of complex pipelines against the Gaia data archive;
            \item Flexible so that developers can use tools and libraries not yet released;
            \item Easy to use, as AVIs will not be developed if the process is too complex.
        \end{enumerate}
        At a high level, the AVI framework is implemented as a web application using the Django web framework~\cite{DjangoProject}. 
        The AVI created by developers forms one component or ``application'' which is run within the AVI framework.
        The framework handles pipeline management, GAVIP communications, and authentication with GAVIP. 
        In addition, the framework supports standalone operation for development purposes. 
        See section \ref{ssub:avi_structure} for detail on the AVI structure.
    
    \subsubsection{AVI development} 
    \label{ssub:avi_development}
        AVI development involves developing an AVI locally, and submitting it for deployment within GAVIP.
        This is achieved by running AVIs (Docker containers) in `standalone' mode on the developers machine. 
        The container images used within GAVIP are publicly available for download, and allow developers to replicate the expected AVI environment with certainty.
        While running in this mode the AVI mounts code which is initially cloned from one of many example AVIs~\cite{ExampleAvis} provided with GAVIP.
        Once the developer is satisfied with their AVI, they simply save the changes and upload their AVI code to the platform (the AVI
        container used locally is not required).
        Further detail on the AVI development process is available in section \ref{ssub:impl_avi_development}.
    
    \subsubsection{Resource management} 
    \label{ssub:resource_management}
        AVIs are designed to run at a large archive in excess of 1 PB in size; as such it is expected that AVIs may often require
        significant resources to operate.
        In addition, AVI resources must be isolated so that they do not impact other AVIs, or indeed the platform, and cannot gain undesired
        access to another users data-products.
        GAVIP must also support hundreds of simultaneous users.
        These requirements provide a significant challenge for any platform, but due to the requirements of GAVIP, there is another layer to the complexity of resource management. AVIs provide a web interface, and serving hundreds of requests is a simple task for any web application. However, as AVIs must only be able to interact with data from the `owner' (the user who wishes to interact with the AVI), they must be provided for one unique user at a time.
        GAVIP satisfies these requirements by creating and running AVIs on demand, and again leveraging the ability for AVIs to operate in different modes. 
        The ability to create and run AVIs on demand is made possible by the underlying use of containers rather than virtual machines. 

        As described previously in this section, AVIs separate interfaces from pipelines in the interest of separating the demands of system
        resources.
        In the case of AVI web interfaces, the AVI is started on demand by a user selecting an AVI from a catalogue. The AVI is provisioned with hundreds of megabytes of RAM, and destroyed when not in use (determined by the user signing out, or session expiring).
        The expected RAM requirements of AVI pipelines will generally surpass the hundreds of megabytes with which AVI interfaces are
        executed.
        Consequently, when a user attempts to invoke a pipeline, the AVI framework automatically dispatches a job request (referred to as an
        AVI task).
        AVI containers are started in order to process these job requests, and are dynamically allocated RAM depending on the unique requirements of each job. 
        This allows AVI developers to, if they wish, provide an algorithm to determine the RAM to be allocated for a particular job, or suggest a RAM allocation to a user, or even allow the user to specify it directly. 
        Considering the context of AVIs and the freedom afforded to developers, this is a significant feature of the platform; it enables
        developers to write arbitrarily complex processes as pipelines and dynamically request the resources to run them, all supported within shared infrastructure.

        To ensure that GAVIP resources are not over-allocated, AVI pipelines are not started immediately; rather they are managed by a
        scheduler.
        The scheduler monitors various metrics of the platform including GAVIP resources, job submission time, jobs submitted per user, and begins an AVI pipeline when appropriate. 
        An AVI pipeline is executed in a separate container to the AVI interface, but they are structured identically; the most significant differences are resource allocations and the processes they are running.
        Ultimately, the solution to resource management within GAVIP is to allow AVIs to run in different modes, run complex processing
        tasks with dynamically allocated resources, and defer those tasks until possible for safe execution.
            

\section{ARCHITECTURE} 
\label{sec:architecture}
    Generally, GAVIP is implemented in Python, apart from Consumer Off The Shelf (COTS) components such as RabbitMQ~\citep{RabbitMQ}. 
All background processes and web interfaces are Python-based. 
Docker is used as GAVIP's container system, with Docker Swarm employed to provide a scalable Docker system across multiple servers. 

\subsection{High-Level Architecture} 
\label{sub:high_level_architecture}
    The \textbf{GAVIP platform} consists of four high-level elements that are listed below with a brief description of their purpose. 
    These elements are illustrated in Figure~\ref{fig:highlevel_components}. 
    
    \begin{enumerate}
        \item \textbf{Portal interfaces:} These allows users to interact with the platform, and access AVIs.
        \begin{enumerate}
          \item \textbf{Graphical user interaction:} Users are able to interact with the GAVIP platform through a graphical web interface rendered in a web-browser.
          \item \textbf{Programmatic user interaction:} The API provided by the portal is generally exercised through the graphical web
          interface. However, GAVIP was designed to support programmatic interaction, allowing the platform to be integrated with other
          scripts or existing tools. As such the API is RESTful, and can already be easily integrated with tools such as the provided GAVIP
          client.
        \end{enumerate}
        \item \textbf{AVI Infrastructure:} Hosts and controls AVI containers.
        \begin{enumerate}
          \item The AVI infrastructure is remotely commanded by the Portal component. It performs operations such as creating, querying,
          updating, and deleting the Docker containers which are used to execute AVIs.
        \end{enumerate}
        \item \textbf{AVI:} user-defined algorithms and interfaces hosted in Docker containers. The different modes of AVI operation are outlined below with a description (previously discussed in section \ref{ssub:resource_management}).
        \begin{enumerate}
          \item \textbf{Standalone:} The AVI operates on a developer's machine providing limited functionality within a 
          replica of the deployed AVI environment. Resources are not restricted. Any jobs triggered through the AVI are executed
          asynchronously within the same container.
          \item \textbf{AVI Interface deployment:} The AVI is deployed within the AVI infrastructure of GAVIP. It is allocated a few hundred
          megabytes of RAM ($\approx$200MB) for operation. Any jobs triggered through the AVI are queued and scheduled in GAVIP.
          \item \textbf{AVI Pipeline deployment:} The AVI is deployed within the AVI infrastructure of GAVIP. It is deployed with a configurable allocation of resources. It does not provide any interface, and is configured to run a single task; the container is destroyed when the task is completed. 
        \end{enumerate}
        \item \textbf{Supporting Systems:} miscellaneous components, employed by the core functionality of the system.
        \begin{enumerate}
          \item The supporting systems consist of miscellaneous software that may be integrated with other componentsate, such as the
          message bus and monitoring service.
        \end{enumerate}
    \end{enumerate}

\subsection{Component Level Architecture} 
\label{sub:component_level_architecture}
\begin{figure}[H]
        \centering
        \includegraphics[width=0.7\textwidth]{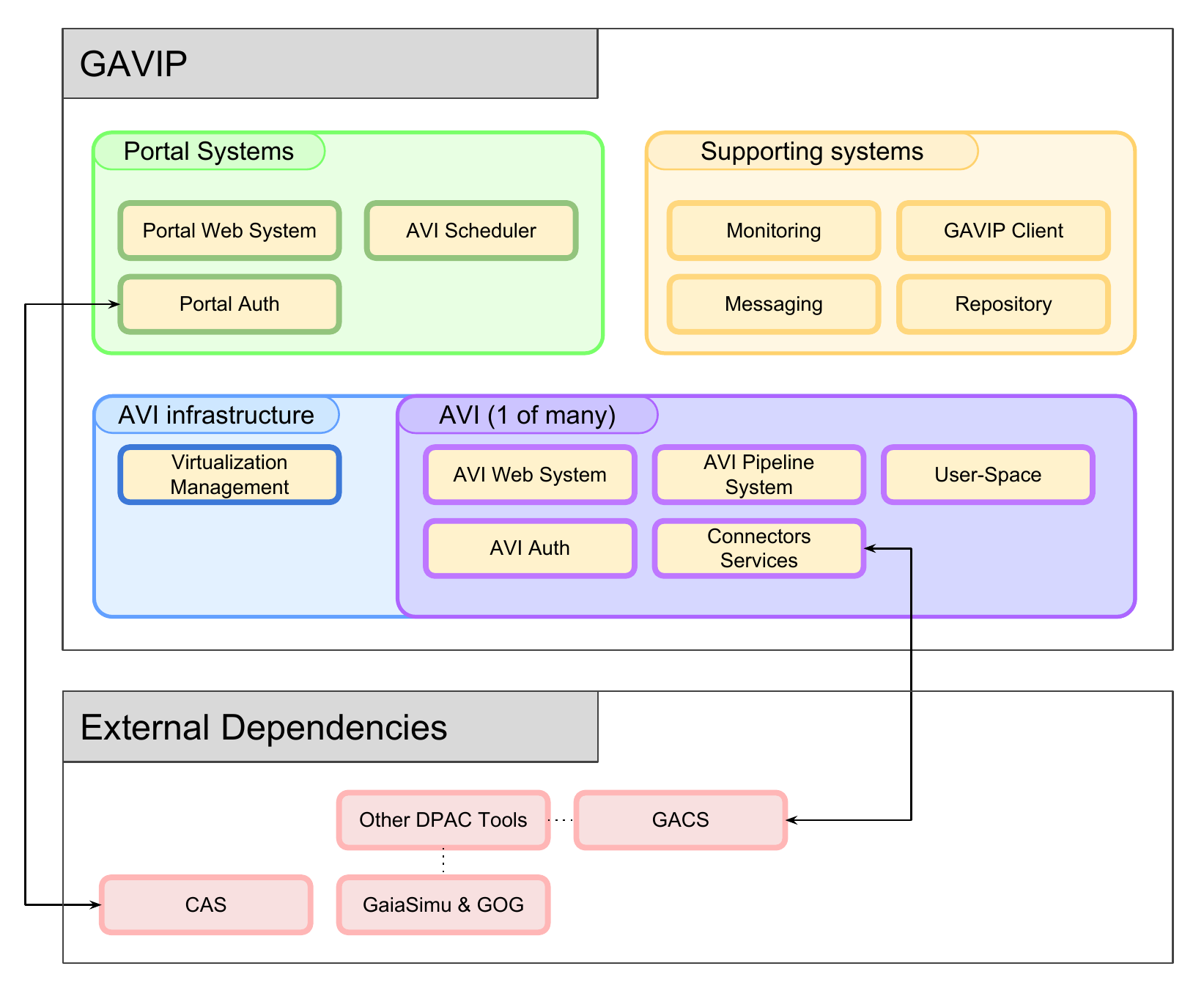}
        \caption{GAVIP component overview showing the 13 main components and their parent high-level elements.}
        \label{fig:component_overview}
    \end{figure}

    The four high-level elements of GAVIP can be further sub-divided into 13 components, listed below with a short description.  An illustration of these 13 components is provided in Figure~\ref{fig:component_overview}.

    \subsubsection{Portal Web System} 
    \label{ssub:portal_web_system}
        The web system provides a RESTful API with a graphical web-interface for users to interact with the GAVIP platform. It is built
        using Django~\citep{DjangoProject}, the Django Rest Framework~\cite{DjangoRestFramework}, and Swagger~\cite{DjangoRestSwagger}.
        The latter is used to build API documentation from the interfaces defined using the Django Rest Framework. The data generated by
        Swagger is used by the GAVIP client (section \ref{ssub:gavip_client}) to dynamically construct its CLI.
        
    \subsubsection{Portal Authentication/Authorisation} 
    \label{ssub:portal_authentication_authorisation}
        This component handles authentication of users within GAVIP. 
        Authentication is offloaded to a (Central Authentication Service) CAS server by the django-cas plugin. 
        By offloading to a CAS server, passwords need not be stored in the platform, thus reducing security concerns.
        A local CAS server is used during GAVIP development; once deployed in ESAC infrastructure, GAVIP will use ESAC's CAS server. 
    
    \subsubsection{Messaging} 
    \label{ssub:messaging}
        The GAVIP platform consists of many components which each often have associated tasks which must be executed, usually in a
        background process. For example, the virtualization controller has many deployment tasks for different types of containers.
        RabbitMQ is used to provide a message bus for GAVIP components to communicate. 
        Celery is used to provide a powerful interface to RabbitMQ.
        By designing GAVIP around a message bus, complex orchestration tasks are easier to control. For example, an outline of the AVI
        pipeline deployment process is as follows:
        \begin{enumerate}
            \item Preparing the user-space.
            \item Downloading the AVI code.
            \item Preparing a queue on the message bus, which is used to send a single job to the AVI.
            \item Starting the container.
            \item Stopping the container once the job is complete.
        \end{enumerate} 
        Some of these tasks may occur within different servers in the platform. 
    
    \subsubsection{Repository} 
    \label{ssub:repository}
        The code for each AVI is maintained within GAVIP using a Git repository. 
        A developer is allocated an individual repository for each AVI they develop. 
        By default, these repositories are private to encourage development and use of tools without exposing source code, but may optionally be exposed publicly.
    
    \subsubsection{AVI Web System} 
    \label{ssub:avi_web_system}
        Provides the web-interface, built using Django.
        Other AVI components such as the AVI pipeline and AVI authentication systems are integrated through this Django web application. 
        AVI developers provide one (Django) \texttt{app}, a single subdirectory within this Django project that is automatically integrated
        into other components such as the pipeline system (the latter is described in more detail in Section~\ref{ssub:avi_pipeline_system}).
    
    \subsubsection{AVI Pipeline system} 
    \label{ssub:avi_pipeline_system}
        Provides the framework for easily submitting and executing pipelines (long running processes) as asynchronous jobs. 
        Pipelines are built and defined by code using Luigi. 
        Although the pipeline framework is written in Python, the AVI framework supports executing Java or Fortran code, which can be called within the pipeline.
        Using the pipeline system is mostly automatic and requires little development effort:
        \begin{enumerate}
            \item Build a pipeline with the required arguments.
            \item Build a Django model that includes those arguments, extending the \texttt{AviJob} model.
            \item Create a Django view which saves an instance of the model (for example, through a form submission).
        \end{enumerate}
        All queueing and job submission occurs automatically using hooks in the \texttt{save()} functions of the \texttt{AviJob} model. Each
        \texttt{AviJob} automatically includes fields for specifying the resource requirementsfor each
        job.
        Furthermore, each \texttt{AviJob} has an associated pipeline state model which stores the progress and state of each job. The
        pipeline state information is used within the platform to build interfaces such as progress bars.
        The progress of each job is calculated using the total number of required tasks, and number of completed tasks in a pipeline. Luigi
        is used for building pipelines in the AVI, and determines these numbers within the framework. Luigi also provides event hooks which are used for updating the job progress.
    
    \subsubsection{AVI Authentication/Authorisation} 
    \label{ssub:avi_authentication_authorisation}
        AVI authentication uses OAuth 2.0~\cite{rfc16749,OAuth2} to authenticate a user within GAVIP. 
        The platform returns a description of the user and their roles (user profile) to the AVI. 
        The user profile is used for authorisation functionality that permits AVI behaviour to be modified based on user roles within GAVIP. 
        For example, alternative AVI interfaces may be provided to outreach users.
        The platform uses the \texttt{oauth2\_provider}~\cite{DjangoOAuth} plugin for Django. 
    
    \subsubsection{Connectors/Services} 
    \label{ssub:connectors_services}
        Connectors and services are reusable libraries available within the AVI framework.
        `Connectors' provide interfaces to thirdparty systems such as Gaia Archive Core Systems (GACS).
        `Services' are sets of pipeline operations that may be easily integrated in any AVI pipeline, typically using one or more
        connectors to provide common operations.
        One example of a connector currently provides a TAP+ interface. 
        In contrast, a service would use the TAP+ connector to submit, monitor, and download the results of an asynchronous Astronomical
        Data Query Language (ADQL)~\citep{ADQL:2008} job.
    
    \subsubsection{Virtualisation Management} 
    \label{ssub:virtualisation_management}
        Manages autonomous deployment of AVIs, minimizing the need for operator intervention; ultimately this component allows containers to be rapidly started on demand. 
        It uses the Python Docker client (\texttt{docker-py}), and connects to a local port that exposes either the standard Docker engine
        or Docker Swarm.
        It is composed of a set of asynchronous tasks for deploying containers, classes which generate the configuration for different types
        of containers, and Django views which provide an API for interfacing with the controller.
    
    \subsubsection{AVI Scheduler} 
    \label{ssub:avi_scheduler}
        Schedules the execution of AVI pipelines with custom resource allocations to manage system load.
        Each scheduler implements a class containing a function that determines the next AVI task to be executed. 
        Multiple classes with different scheduling algorithms will be developed and can be selected in the GAVIP platform settings.
    
    \subsubsection{User-space} 
    \label{ssub:user_space}
        Provides a persistent storage system for each user.
        The user-space is currently implemented as an NFS volume (though this is subject to change). 
        Alternative methods of user-space storage may be object bases (similar to Amazon S3).
        Docker can be extended to use alternative storage engines through its plugin system.    
        All AVIs started by each user will be configured to mount their unique user-space for reading and writing data products.
        A user-space browser is included in the portal web system that enables users to browse, upload, download, and delete the
        corresponding files.
    
    \subsubsection{Monitoring} 
    \label{ssub:monitoring}
        GAVIP components, including AVIs, are monitored by Zabbix~\cite{Zabbix}, which provides a web dashboard similar to
        Nagios~\cite{Nagios} and Munin~\cite{Munin}.
    
    \subsubsection{GAVIP client} 
    \label{ssub:gavip_client}
        Provides a command line interface to GAVIP, permitting almost all platform operations to be invoked programmatically without the
        need for a web browser.
        Additional commands for AVI development are included in the client that invoke Docker commands on the local machine.
        For example, a \texttt{quickstart} command is provided that downloads, prepares, and starts an example AVI~\cite{ExampleAvis}.
        Although Docker is run on Linux and currently requires a Virtual Machine to run on Windows or MacOS systems, native solutions are being developed which will remove this requirement~\cite{DockerMacWinBeta}.
        The GAVIP client is built using the Python \texttt{click} module. 
        The algorithm which parses the platform API and generates the CLI commands (via data provided by Swagger in the Portal web system) was developed internally. 

\subsection{Feature Implementation} 
\label{sub:technical_implementation}
    \subsubsection{AVI structure} 
    \label{ssub:avi_structure}
\begin{figure}[H]
            \centering
            \includegraphics[width=1\textwidth]{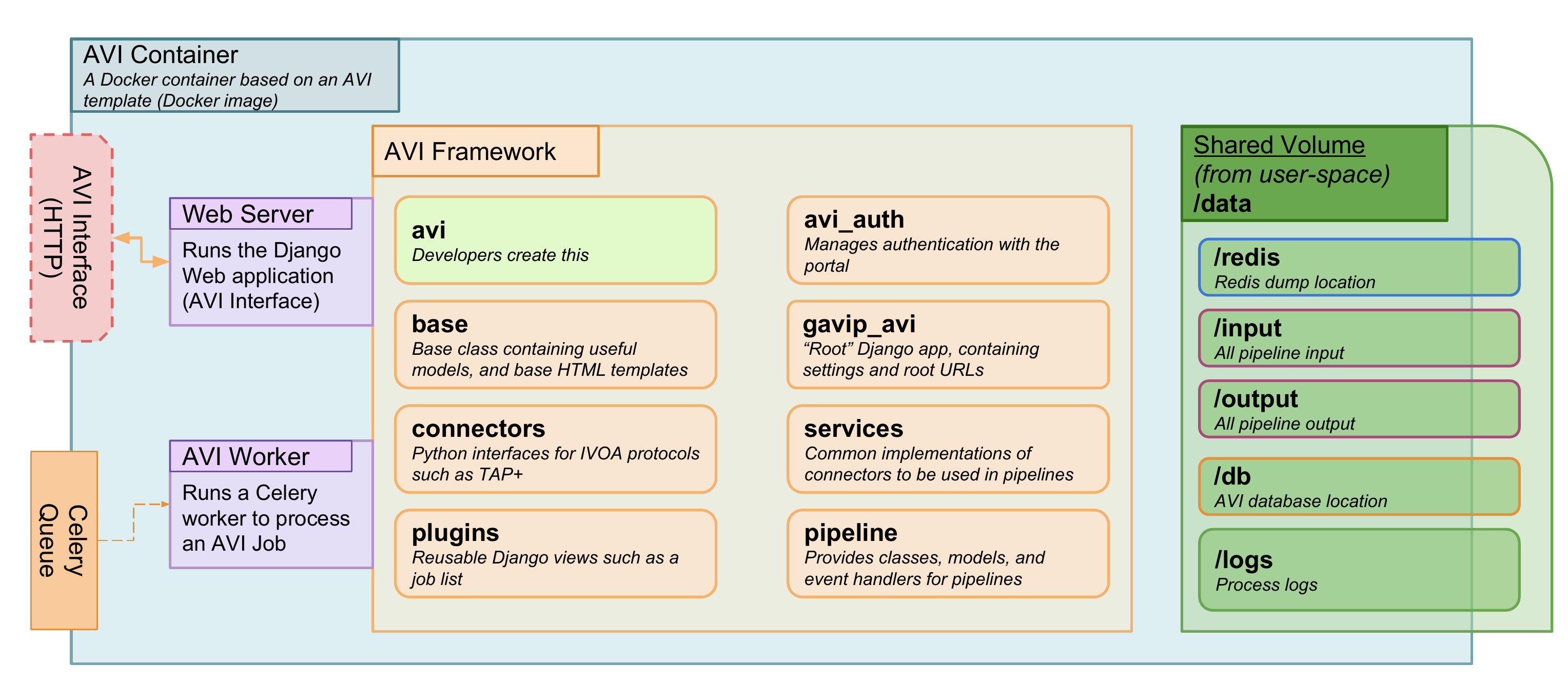}
            \caption{Internal AVI illustration, consisting of the AVI framework executing the \texttt{avi} code provided by a developer.
            The AVI framework is running in a container that is based on one of the provided AVI templates. Two modes of operation are shown
            for the AVI framework (`Web Server' and `AVI Worker'). When the AVI is running a `Web Server', it exposes a HTTP interface to the AVI. Alternatively, when running an `AVI Worker', it processes a single AVI job (an AVI pipeline).}
            \label{fig:avi_component_illustration}
        \end{figure}

        The AVI structure is described here to provide context for subsequent sections.
        An illustration of the AVI structure including the applications within the AVI Framework can be seen in Figure~\ref{fig:avi_component_illustration}.
        The green box labelled ``Shared Volume'' represents the mounted user-space directory, where the AVI stores persistent data-products.
        The two purple boxes labelled ``Web Server'' and ``AVI Worker'' represent two modes of operation for the AVI framework; the former provides a web interface for using the AVI, while the latter processes a single AVI job. 
        AVI jobs are created by users interacting with the interface and submitting jobs to the GAVIP platform, or to an internal worker when running in ``standalone'' mode.
        A description of each of the framework applications is provided in Table~\ref{tab:avi_components}. The AVI framework also provides a
        wide array of installed packages available for use (by including an Anaconda bundle), for example:
        
        \begin{scriptsize}                
            \begin{multicols}{4}
                \begin{itemize}
                    \itemsep0em 
                    \item astroML (v0.3)
                    \item astropy (v1.1.2)
                    \item astroquery (v0.3.1)
                    \item blaze (v0.9.1)
                    \item bokeh (v0.11.1)
                    \item dask (v0.8.1)
                    \item Jinja2 (v2.8)
                    \item lxml (v3.6.0)
                    \item matplotlib (v1.5.1)
                    \item numba (v0.24.0)
                    \item numexpr (v2.5)
                    \item numpy (v1.10.4)
                    \item pandas (v0.18.0)
                    \item pandas-profiling (v1.0.0a2)
                    \item psycopg2 (v2.6.1)
                    \item pyfits (v3.4)
                    \item scikit-image (v0.12.3)
                    \item scikit-learn (v0.17.1)
                    \item scipy (v0.17.0)
                    \item seaborn (v0.7.0)
                    \item statsmodels (v0.6.1)
                \end{itemize}
            \end{multicols}
        \end{scriptsize}

        \begin{table}[H]
            \centering
            \begin{tabular}{|p{4cm}|p{12cm}|}
                \hline
                \textbf{Application} & \textbf{Description} \\ \hline
                \hline
                \texttt{avi} & Code written by the AVI developer. It will typically include model definitions, Django views, URLs, and HTML
                templates. \\ \hline 
                \texttt{avi\_auth} & Handles authenticating the AVI within the platform. OAuth is used as the AVI cannot access the GAVIP
                database. GAVIP returns a user profile object which contains details of the user. \\ \hline 
                \texttt{base} & Contains base HTML for providing a consistent look and feel to the AVI. \\ \hline
                \texttt{gavip\_avi} & The root Django application which defines the installed applications, root URL structure, and configuration of the AVI. \\ \hline \texttt{connectors} & Python interfaces for IVOA protocols. See Section \ref{ssub:connectors_services}. \\ \hline
                \texttt{services} & Pipeline tasks which implement some usage of a connector. For example, a `GACS' service is provided
                which uses the TAP+ connector to manage a job from GACS. See Section \ref{ssub:connectors_services}. \\ \hline
                \texttt{plugins} & Provides reusable Django views to reduce web development effort required by AVI developers. \\ \hline \texttt{pipeline} & Includes the classes, event handlers and tasks required for managing AVI jobs.  \\ \hline
            \end{tabular}
            \caption{AVI Framework Applications.}
            \label{tab:avi_components}
        \end{table}
         

    \subsubsection{AVI pipelines} 
    \label{ssub:avi_pipelines}
\begin{figure}[H]
            \centering
            \includegraphics[width=0.7\textwidth]{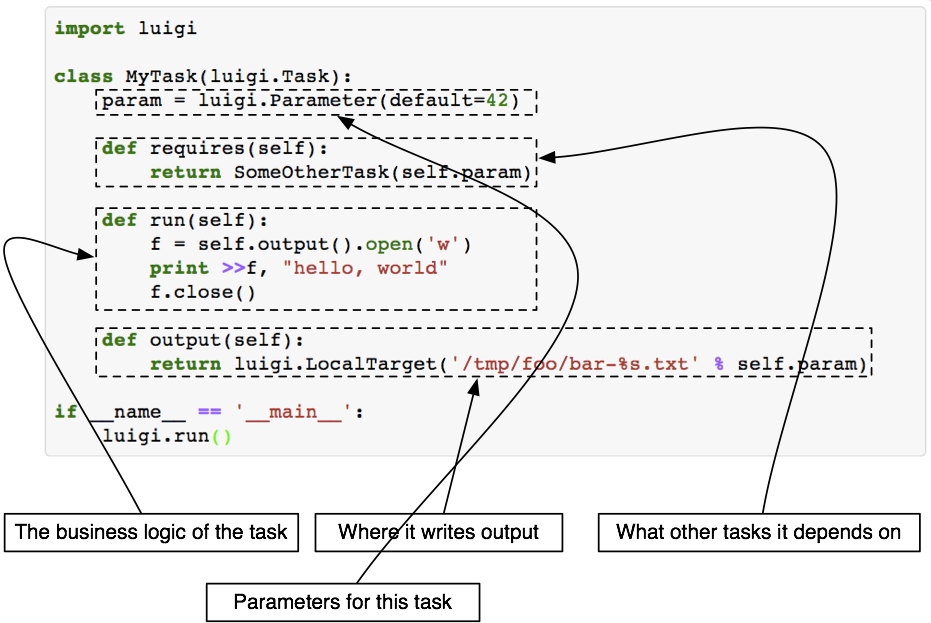}
            \caption{A sample Luigi task showing how the pipeline structure is defined by code by explicitly defining parameters and using 
            \texttt{requires()}, \texttt{run()} and \texttt{output()} functions. \textbf{Source:} Luigi documentation \cite{LuigiRTD}.}
            \label{fig:sample_luigi_task}
        \end{figure}

\begin{figure}[H]
            \centering
            \fbox{\includegraphics[width=0.7\textwidth]{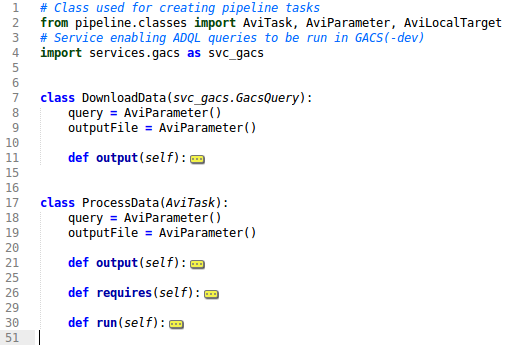}}
            \vskip +2.0em
            \caption{A screenshot of a sample pipeline showing two tasks \texttt{DownloadData} and \texttt{ProcessData}. The classes
            \texttt{AviTask}, \texttt{AviParameter}, and \texttt{AviLocalTarget} are all extensions of Luigi classes. The
            \texttt{DownloadData} task extends the service class \texttt{services.gacs.GacsQuery}, which itself is an
            \texttt{AviTask}. In this example, \texttt{GacsQuery} implements the
            \texttt{run()} function that will be used during the execution of the \texttt{DownloadData} task, thus demonstrating how
            services can be used to include common functionality in AVI pipelines.}
            \label{fig:sample_avi_pipeline}
        \end{figure}

        AVI pipelines are used to execute all long-running processes within an AVI.
        Pipelines are implemented using a combination of Luigi~\cite{Luigi} and Celery~\cite{Celery}. 
        Luigi allows pipelines to be defined by code. Celery is used to manage queueing of asynchronous jobs.
        A sample Luigi task can be seen in Figure~\ref{fig:sample_luigi_task}, with a short sample AVI pipeline shown in Figure~\ref{fig:sample_avi_pipeline} (note that the code is collapsed in these figures for brevity).
        Each task of an AVI pipeline (e.g. \texttt{DownloadData} and \texttt{ProcessData} in Figure~\ref{fig:sample_avi_pipeline}) are
        inspected before execution to see if the output defined by the step already exists.
        If this is true, the task is skipped. 
        This allows developers to easily prevent unnecessary computation (for example, a task output file may be constructed using a hash of
        the corresponding input parameters).

        Once a pipeline is created in the AVI framework, a model (Django model) must be created to store the associated arguments for the pipeline. 
        The job request handled by Celery only requires the primary key of the model instance rather than all the arguments to the
        pipeline. This simplifies the Celery message content, and bypasses the need for serializing complex objects such as input files.
        The model created by the developer extends a model class provided by the framework called \texttt{AviJob}.
        The \texttt{AviJob} model provides logic for creating the AVI job request and submitting it to the portal. Once the
        model is saved (e.g. upon a form submission in the AVI), a task is submitted for processing.
        As described in Section~\ref{ssub:impl_resource_management}, AVIs have different modes of operation so that resources can be allocated when required. 
        AVIs may run on a developers machine in ``standalone'' mode or may run within the GAVIP platform (either to provide a web interface or process a single AVI job).
        No further development is required by developers to support these different modes of operation; decoupling of AVI behaviour happens
        automatically, and is part of the logic provided by the \texttt{AviJob} model.


    \subsubsection{AVI development} 
    \label{ssub:impl_avi_development}
        The AVI development process is described in this section, with prerequisite information regarding AVI images and development
        environment provided beforehand.

        \paragraph{AVI template} 
        \label{par:avi_template_creation}
            Before an AVI is developed, a GAVIP operator must make one or more ``AVI templates'' available. 
            These are Docker images built for the platform, and provide an image from which AVI containers are created.
            A list of some of the available images can be found in Table~\ref{tab:avi_templates}. 
            These images may be downloaded from \url{https://repositories.gavip.science} using the associated image identifiers, which 
            are distributed in the GAVIP Developer manuals that are provided with each release of the platform.
            Additional AVI templates will be provided at a later date to support the use of additional languages such as Fortran.

            \begin{table}[H]
                \centering
                \begin{tabular}{|p{4cm}|p{12cm}|}
                    \hline
                    \textbf{Template} & \textbf{Description} \\ \hline
                    \hline
                    Python AVI & It includes all components required to execute either an AVI web interface or AVI pipeline. \\ \hline
                    Java (7) AVI & An extension of the Python AVI template, which permits the execution of AVIs that have Java (7) dependencies. \\ \hline
                    Java (8) AVI & An extension of the Python AVI template, which permits the execution of AVIs that have Java (8) dependencies. \\ \hline
                \end{tabular}
                \caption{AVI Templates.}
                \label{tab:avi_templates}
            \end{table}


        \paragraph{AVI project} 
        \label{par:avi_project}
            AVIs are created in GAVIP using an ``AVI template'' and a particular release version of an ``AVI project''.
            An \textbf{AVI project} is a record consisting of a particular Git repository (which hosts the AVI code), and some metadata such as AVI title and author.
            An \textbf{AVI release} is a record which provides a snapshot of the AVI project repository. 
            The record consists of a Git hash (a unique identifier for each change to the repository), a version number, and a flag indicating whether the release is to be public or not.
            If the release is not made public, it will only be made available to the AVI developer rather than all users of the platform.

        \paragraph{Development environment} 
        \label{par:development_environment}
            A developer guide is provided by GAVIP to help users create AVIs. 
            The guide includes a list of prerequisite requirements including particular packages, with Docker and the GAVIP Client being two
            key examples of the latter. Docker is used to run the AVI in `standalone' mode on the developers machine during
            development, while the GAVIP client automates some of the interactions with both the GAVIP platform and Docker for
            the sake of convenience.

\begin{figure}[!ht]
                \centering
                \includegraphics[width=1\textwidth]{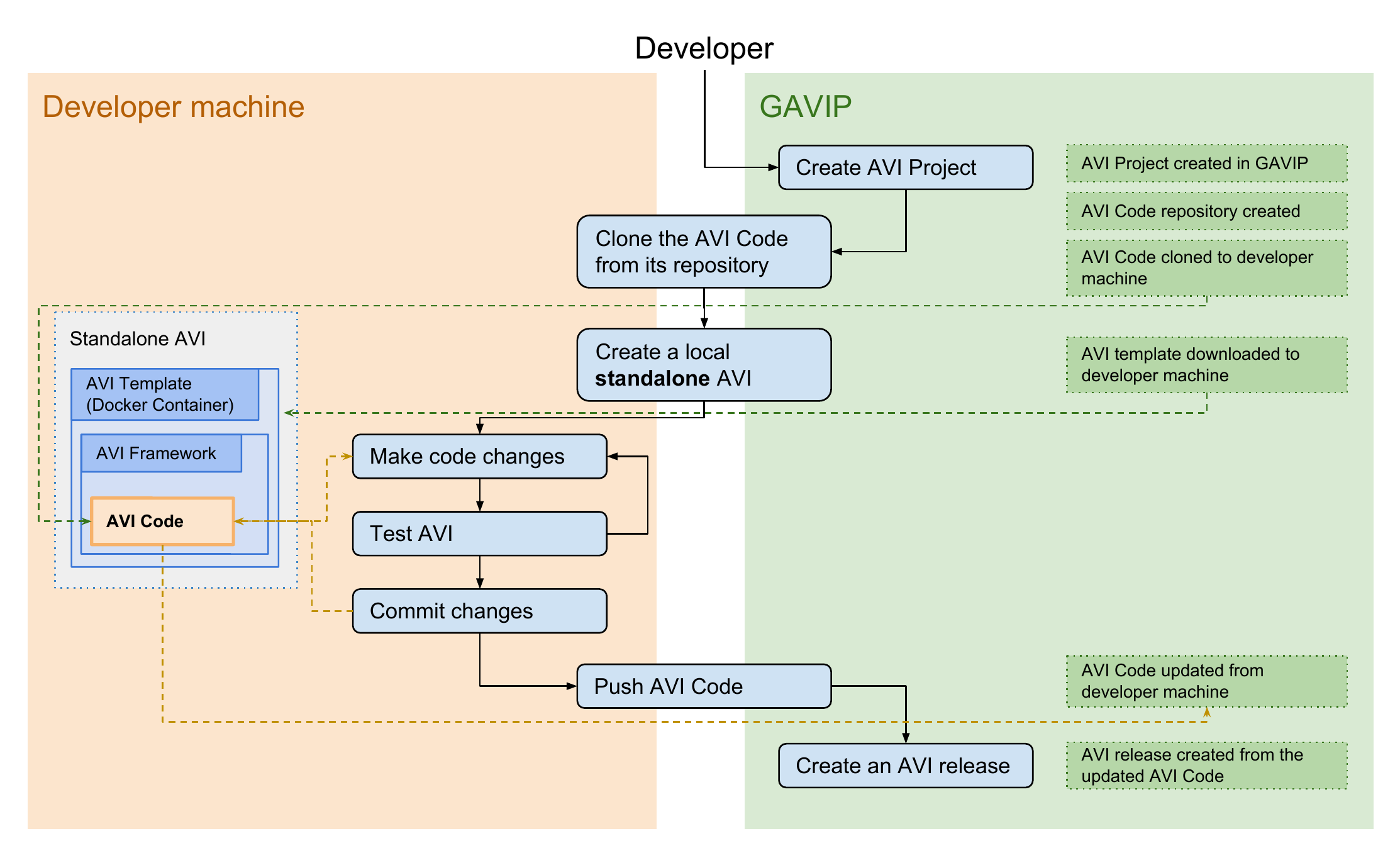}
                \caption{The AVI development process, indicating the steps occurring on the developer machine (left),
                and steps occurring in the GAVIP platform (right). The steps themselves are shown in the blue rounded boxes, with arrows
                indicating the flow of operations. Dotted green boxes on the right provide further details. A conceptual
                representation of a Standalone AVI is shown within the developer machine, with dotted lines used to identify operations
                that interact with the AVI.}
                \label{fig:develop_process}
            \end{figure}

        \paragraph{Development process} 
        \label{par:development_process}
            AVIs will typically begin with a concept algorithm to form the basis of the AVI.
            Developers are encouraged to use a Jupyter notebook (hosted within GAVIP) to prototype an algorithm before AVI development.
            For more information on the execution of Jupyter notebooks in the platform, refer to section \ref{ssub:jupyter_notebook}.
            AVI developers will follow the AVI development guide provided by GAVIP. 
            As part of this, they set up any prerequisites such as Docker and the GAVIP client. 
            The development process is outlined in Figure~\ref{fig:develop_process}.
            When a developer creates a public AVI release, the AVI is immediately available within the GAVIP platform for other users to use.


    \subsubsection{Jupyter notebooks} 
    \label{ssub:jupyter_notebook}
        Jupyter notebooks can be executed within GAVIP by any user.
        They are started in an isolated container built from the Python AVI template; this permits a notebook to import and use features
        within the AVI framework (e.g. the TAP+ connector).
        The notebooks are saved in the user-space, and can be downloaded using the user-space browser.

    \subsubsection{Dashboards} 
    \label{ssub:dashboards}
        The GAVIP portal provides a unique interface that permits multiple AVIs to be simultaneously viewed and dynamically
        rearranged and resized by the user.
        The dashboard is implemented using Gridster~\cite{GridsterJS} and HTML Iframes. 
        Gridster is used as it enables HTML \texttt{$<$div$>$} tags to be resized and repositioned within a web page. HTML Iframes
        permit AVI views to be loaded within these \texttt{$<$div$>$} tags.
        A SAMP Hub is available for each user of GAVIP, and is intended to provide a means of
        communication between multiple AVIs.
        Other tools such as TopCat may be connected to the GAVIP SAMP hub, which would require the use of a proxy using the SAMP web
        profile.
        The AVI framework includes a ``plugin'' which can be used in any of the AVI web pages to initialize and expose a SAMPJS connector to the user's SAMP hub.

        Users may add and remove AVI views from their dashboard as they wish. 
        If an AVI is not running, and is required by the dashboard, it is created and started (this process takes the order of seconds).
        A catalogue of AVI views will be provided for the user when they wish to add a new view.
        It is expected that these views will be identified when an AVI is submitted to GAVIP through a naming convention (e.g.
        \texttt{dashboard\_3d\_star\_plot()}); this is yet to be finalized.
        Each user may save arrangements of AVI views within the dashboard as ``dashboard presets''.
        Developers can also create ``dashboard presets'', and these are made available to all users.
        These presets are persisted as records within the GAVIP database. 

    \subsubsection{Resource management} 
    \label{ssub:impl_resource_management}
        
        GAVIP resource management involves recording the following:
        \begin{itemize}
            \item Resources required for different processes.
            \begin{itemize}
                \item For example, the resources required by a container running an AVI interface or pipeline.
             \end{itemize} 
            \item Total resources currently allocated per user.
            \item Maximum allocation of resources per user at any instant.
         \end{itemize}
        
        Each user is assigned a limit to the amount of resources they can have allocated at any point in time; this is known as a
        \textbf{resource pool}.
        The resource pool is used to inform the user, and help the user consider that certain operations require resources in the platform.
        All containers used within GAVIP extend a \texttt{GavipContainer} class, which has attributes used for recording the resources to be
        allocated to that container.
        For most containers, these values are read from a configuration file. However, for containers running AVI pipelines, they are read
        from the associated AVI job (sent by the AVI).
        The \texttt{AviJob} model provided in the AVI framework contains fields for storing the resources to be allocated when running the
        pipeline.
        As this is stored in the \texttt{AviJob} model, it may be modified by the developer or by the user as appropriate, depending
        purely on the algorithm being used and structure of the pipeline. Occasionally, a resource-intensive portion of the
        pipeline is skipped as the output file has already been created (see Section~\ref{ssub:avi_pipelines}).

        When a container is being started (e.g. an AVI pipeline container, or AVI interface container) the resources required are checked against the free resources available (determined by subtracting the allocated resources from the resource pool).
        Every container used in GAVIP has an associated model that is used to track all container deployments. 
        Whenever the container model is saved, the allocated resources are recalculated.

    \subsubsection{GAVIP client} 
    \label{ssub:impl_gavip_client}
        The GAVIP client provides a command line interface to the GAVIP platform, and assists AVI development.
        It is built using Click~\cite{Click}. 
        The commands in the client are dynamically generated by parsing API documentation hosted by the GAVIP platform.
        This API documentation is provided by Django-Rest-Swagger~\cite{DjangoRestSwagger} working with the Django Rest Framework~\cite{DjangoRestFramework} and is hosted within the portal.
        Django-Rest-Swagger is a Django plugin that provides Swagger API documentation. As part of this process, a hierarchical data
        representation of the API is made available which details:
        \begin{itemize}
            \item The URL of a particular view.
            \item A function nickname.
            \item Documentation from the function docstring.
            \item Input parameter details (including parameter type where possible).
            \item Output parameter details.
        \end{itemize}
        When a user logs in to the GAVIP platform using the client, this API is read and used to dynamically generate and decorate functions which interact with the GAVIP platform.
        The Click module includes an autocomplete function that works with these generated functions.
        A demo of API generation can be viewed online\footnote{\url{https://asciinema.org/a/44809}}.

\section{CONCLUSIONS AND FUTURE WORK} 
\label{sec:conclusions}
    The GAVIP platform is designed to enable communities to incorporate existing codes and develop an expanding repository of tools to engage
with large data-archives (the Gaia data archive in this case) without requiring the archive to be downloaded.
There are many related and driving requirements behind the design of GAVIP, some of which have been refined in Table~\ref{tab:gavip_requirements}.
GAVIP is required to be highly automated, able to run multiple AVIs in isolation, and dynamically control the demands on system resources.
By using containers to host AVIs, we have maximized the isolation of AVIs while minimising the resources required to do so. 
Designing AVIs such that they have different modes of operation allows the demands on system resources to be decoupled and managed separately.
Combined, these two features allow the platform to meet automation, isolation, scalability, and load control requirements.
The AVI framework is designed to meet the requirements of AVI flexibility and web-based interaction by supporting developers in creating
complex pipelines with few restrictions, and interfaces using the latest web-technologies, as well as providing an expanding range of
libraries and packaged utilities such as ``Connectors'' and ``Services'' (see Section~\ref{ssub:connectors_services}). More generally, the
platform also provides a mechanism for addressing the contemporary issue of scientific experiment reproducibility. In addition, the proposed
design and implementation mean that it is suitable for use in domains beyond astronomy.

The Gaia archive data will be made available in multiple releases, where early releases will consist of restricted subsets of validated
data with the individual epoch observations and transits appearing only in the final release. (an overview of the release scenario can
be found online \footnote{\url{http://www.cosmos.esa.int/web/gaia/release}}).
The GAVIP platform is scheduled for early deployment toward the end of 2016, potentially overlapping with the first Gaia data release.
Final acceptance of GAVIP is scheduled for Q3 of 2017, potentially overlapping with the second of the five expected Gaia data releases.
Three exemplar AVIs (GAVITA, GAVISC, and GAVIDAV) are being developed in tandem with GAVIP to be deployed within the platform prior to final acceptance. 
The first of these AVIs, GAVITA will provide added value in the domain of transient analysis (specifically, supernovae analysis).
GAVITA is developed by Parameter Space in tandem with GAVIP; this helps establish an internal feedback loop of requirements and features of the platform.
GAVISC will model and analyse asteroids detected by Gaia using a combination of Python and Fortran.
GAVIDAV will provide a suite of visualisation tools which can be used on the Gaia archive, as well as data-products created by users in the platform.
These AVIs will be installed and available within GAVIP before final acceptance of the platform.

As part of GAVIP development and design, additional features of the platform have been identified that will be researched in the
future, for example, the compute power available to AVIs.
Currently, AVI pipelines run solely within a Docker container using resources from the host system, rather than being run on a compute
cluster such as Apache Hadoop~\citep{White:2009:HDG:1717298}.
Although not all AVIs will require a compute cluster, access to such a facility will enhance the potential impact of the platform.
The platform is technically capable of exploiting a compute cluster, as the software used to define pipelines with the AVI
framework (Luigi) already supports the integration of separate Apache Hadoop or Apache Spark~\citep{Zaharia:2010:SCC:1863103.1863113}
clusters. 
Other GAVIP modifications for cluster integration include extension of the existing resource
pool implementation to interact with the cluster's own resource manager, for example Apache Hadoop
YARN~\cite{Vavilapalli:2013:AHY:2523616.2523633}, and managing user behaviour to prevent unnecessary usage of the cluster for trivial
jobs.
It is intended that compute cluster integration will be formally included in future releases of GAVIP.


\appendix    

\acknowledgments     
 
This work is funded by the European Space Agency under contract number 4000112862/14/NL/JD. David Lynn acknowledges funding from the Irish
Research Council and Parameter Space Ltd. under the Enterprise Partnership Scheme (grant EPSPG/2015/26).

\bibliography{gavip}   

\begin{thebibliography}{10}

\bibitem{2001A&A...369..339P}
{Perryman}, M.~A.~C., {de Boer}, K.~S., {Gilmore}, G., et~al., ``{GAIA:
  Composition, formation and evolution of the Galaxy},'' {\em Astronomy and
  Astrophysics}~{\bf 369},  339--363 (Apr. 2001).

\bibitem{2009PASP..121.1395L}
{Law}, N.~M., {Kulkarni}, S.~R., and {Dekany}, R.~G.~o., ``{The Palomar
  Transient Factory: System Overview, Performance, and First Results},'' {\em
  Publications Of The Astronomical Society Of The Pacific}~{\bf 121},
  1395--1408 (Dec. 2009).

\bibitem{2010SPIE.7733E..0EK}
{Kaiser}, N. et~al., ``{The Pan-STARRS wide-field optical/NIR imaging
  survey},'' in [{\em Society of Photo-Optical Instrumentation Engineers (SPIE)
  Conference Series}{\nolinebreak\hspace{0.1em}]},  {\em Society of
  Photo-Optical Instrumentation Engineers (SPIE) Conference Series} {\bf 7733},
   0 (July 2010).

\bibitem{ivezic2008lsst}
Ivezic, Z., Tyson, J., Acosta, E., Allsman, R., Anderson, S., Andrew, J.,
  Angel, R., Axelrod, T., Barr, J., Becker, A., et~al., ``{LSST: from science
  drivers to reference design and anticipated data products},'' {\em arXiv
  preprint arXiv:0805.2366}  (2008).

\bibitem{Gaia}
{European Space Agency}, ``{The Detectability of Orphan Afterglows},'' {\em ESA
  Special Publication}~{\bf SP-1323} (2012).

\bibitem{2003MNRAS.341..569B}
{Belokurov}, V.~A. and {Evans}, N.~W., ``{Supernovae with `super-Hipparcos'},''
  {\em Monthly Notices of the Royal Astronomical Society}~{\bf 341},  569--576
  (May 2003).

\bibitem{LuriOverviewCatalogueGOG2014}
{Luri}, X., {Palmer}, M., {Arenou}, F., et~al., ``{Overview and stellar
  statistics of the expected Gaia Catalogue using the Gaia Object Generator},''
  {\em Astronomy \& Astrophysics}~{\bf 566},  A119 (June 2014).
\newblock \url{http://adsabs.harvard.edu/abs/2014A%26A...566A.119L}.

\bibitem{RobinGUMS2012}
{Robin}, A.~C., {Luri}, X., and {Reyl{\'e}}, C.~o., ``{Gaia Universe model
  snapshot. A statistical analysis of the expected contents of the Gaia
  catalogue},'' {\em Astronomy \& Astrophysics}~{\bf 543},  A100 (July 2012).
\newblock \url{http://adsabs.harvard.edu/abs/2012A%26A...543A.100R}.

\bibitem{2012Ap&SS.341..163A}
{Altavilla}, G., {Botticella}, M.~T., {Cappellaro}, E., and {Turatto}, M.,
  ``{Supernovae and Gaia},'' {\em Astrophysics and Space Science}~{\bf 341},
  163--178 (Sept. 2012).

\bibitem{Merkel:2014:DLL:2600239.2600241}
Merkel, D., ``{Docker: Lightweight Linux Containers for Consistent Development
  and Deployment},'' {\em Linux Journal}~{\bf 2014} (Mar. 2014).

\bibitem{DockerImages}
Docker, ``{What is Docker?}.'' \url{https://www.docker.com/what-docker}.

\bibitem{numpy2011}
van~der Walt, S., Colbert, S.~C., and Varoquaux, G., ``{The NumPy Array: A
  Structure for Efficient Numerical Computation},'' {\em Computing in Science
  Engineering}~{\bf 13},  22--30 (March 2011).

\bibitem{scipy2001}
Jones, E., Oliphant, T., Peterson, P., et~al., ``{SciPy}: Open source
  scientific tools for {Python}.''

\bibitem{GoogleDatalab}
Google, ``{Google Datalab}.'' \url{https://cloud.google.com/datalab/}.

\bibitem{MyBinder}
binder project, ``Binder.'' \url{http://mybinder.org/}.

\bibitem{Jupyter}
{Project Jupyter}, ``{Jupyter Notebook}.'' \url{http://jupyter.org/}.

\bibitem{CADC}
{National Research Council of Canada (NRC)}, ``{Canadian Astronomy Data
  Centre}.'' \url{http://www.cadc-ccda.hia-iha.nrc-cnrc.gc.ca/en/}.

\bibitem{SciServer}
{IDIES, John Hopkins University}, ``{SciServer - A collaborative research
  environment for large-scale data-driven science}.''
  \url{http://www.sciserver.org/}.

\bibitem{SciServerTools}
{IDIES, John Hopkins University}, ``{A collaborative research environment for
  large-scale data-driven science}.'' \url{http://www.sciserver.org/tools/}.

\bibitem{DjangoProject}
{Django Software Foundation}, ``{Django}.'' \url{https://djangoproject.com}.

\bibitem{ExampleAvis}
{Parameter Space Ltd.}, ``{Example AVIs for the GAVIP platform}.''
  \url{https://github.com/parameterspace-ie/example-avis}.

\bibitem{RabbitMQ}
Pivotal, ``{RabbitMQ}.'' \url{https://www.rabbitmq.com/}.

\bibitem{DjangoRestFramework}
Christie, T., ``{Django REST Framework}.''
  \url{http://www.django-rest-framework.org/}.

\bibitem{DjangoRestSwagger}
Gibbons, M., ``{Django REST Swagger}.''
  \url{http://django-rest-swagger.readthedocs.io/en/latest/}.

\bibitem{rfc16749}
Hardt, D., ``{The OAuth 2.0 Authorization Framework},'' {RFC} 6749 (October
  2012).

\bibitem{OAuth2}
``{OAuth 2.0}.'' \url{http://oauth.net/2/}.

\bibitem{DjangoOAuth}
{Evonove}, ``{Django OAuth Toolkit}.''
  \url{https://github.com/evonove/django-oauth-toolkit}.

\bibitem{ADQL:2008}
Ortiz, I., Lusted, J., Dowler, P., Szalay, A., Shirasaki, Y.,
  Nieto-Santisteban, M.~A., Ohishi, M., O'Mullane, W., Osuna, P., {VOQL-TEG},
  and {the VOQL Working Group}, ``{IVOA Astronomical Data Query Language},''
  International Virtual Observatory Alliance (May 2008).

\bibitem{Zabbix}
{Zabbix LLC}, ``Zabbix.'' \url{http://www.zabbix.com/}.

\bibitem{Nagios}
{Nagios Enterprises}, ``Nagios.'' \url{https://www.nagios.org/}.

\bibitem{Munin}
``Munin.'' \url{http://munin-monitoring.org/}.

\bibitem{DockerMacWinBeta}
Docker, ``{Docker for Mac and Windows beta}.''
  \url{https://blog.docker.com/2016/03/docker-for-mac-windows-beta/}.

\bibitem{Luigi}
Spotify, ``Luigi.'' \url{http://luigi.readthedocs.io/en/stable/index.html}.

\bibitem{Celery}
Solem, A., ``Celery.'' \url{http://www.celeryproject.org/}.

\bibitem{LuigiRTD}
Spotify, ``{Luigi Tasks}.''
  \url{http://luigi.readthedocs.io/en/stable/tasks.html}.

\bibitem{GridsterJS}
{Ducksboard (New Relic)}, ``{GridsterJS}.''
  \url{https://github.com/ducksboard/gridster.js}.

\bibitem{Click}
Ronacher, A., ``Click.'' \url{http://click.pocoo.org/5/}.

\bibitem{White:2009:HDG:1717298}
White, T.,  [{\em {Hadoop: The Definitive Guide}}{\nolinebreak\hspace{0.1em}]},
  O'Reilly Media, Inc., 1st~ed. (2009).

\bibitem{Zaharia:2010:SCC:1863103.1863113}
Zaharia, M., Chowdhury, M., Franklin, M.~J., Shenker, S., and Stoica, I.,
  ``{Spark: Cluster Computing with Working Sets},'' in [{\em Proceedings of the
  2Nd USENIX Conference on Hot Topics in Cloud
  Computing}{\nolinebreak\hspace{0.1em}]},  {\em HotCloud'10},  10--10, USENIX
  Association, Berkeley, CA, USA (2010).

\bibitem{Vavilapalli:2013:AHY:2523616.2523633}
Vavilapalli, V.~K., Murthy, A.~C., Douglas, C., Agarwal, S., Konar, M., Evans,
  R., Graves, T., Lowe, J., Shah, H., Seth, S., Saha, B., Curino, C., O'Malley,
  O., Radia, S., Reed, B., and Baldeschwieler, E., ``{Apache Hadoop YARN: Yet
  Another Resource Negotiator},'' in [{\em Proceedings of the 4th Annual
  Symposium on Cloud Computing}{\nolinebreak\hspace{0.1em}]},  {\em SOCC '13},
  5:1--5:16, ACM, New York, NY, USA (2013).

\end{thebibliography}
\bibliographystyle{spiebib} 

\end{document}